\documentclass[twocolumn,secnumarabic,amssymb, nobibnotes, aps]{revtex4-1}
\usepackage{graphicx}
\usepackage{subfigure}
\usepackage{amsmath}
\usepackage{epstopdf}

\begin{document}

\title[Short Title]{Single-photon Quantum Router by Two Distant Artificial Atoms}

\author{Xin-Yu Chen$^{1}$}
\author{Feng-Yang Zhang$^{1,2}$}
\author{Chong Li$^{1}$}
\email{lichong@dlut.edu.cn}

\affiliation{$^{1}$School of Physics and Optoelectronic Engineering, Dalian University of Technology, Dalian 116024, China}
\affiliation{$^{2}$School of Physics and Materials Engineering, Dalian Nationalities University, Dalian 116600, China}

\begin{abstract}
Nowadays, quantum router is playing a key role in quantum communication and quantum networks. Here we propose a tunable single-photon routing scheme, based on quantum interference, which uses two distant artificial atoms coupling to two transmission lines. Depending on the distance between the two atoms, the collective effect will lead to destructive or constructive interference between the scattered photons. Under standing wave condition, single photon from the incident channel can be perfectly transmitted or redirected into another channel by asymmetric or symmetric reflected phase shift of atoms, respectively. Therefore, we show that our router can be controlled by adjusting the detuning of the atoms and interatomic distance, without any classical field.
\end{abstract}

\maketitle

\section{Introduction}

Efficient quantum network is a very important ingredient for transmitting information quickly over large distances in quantum information \cite{Nat2008-453-1023}. It consists of quantum channels and nodes, which are provided by waveguides and quantum emitters, respectively. Photons are the most suitable carriers for transmitting information due to the fact that they propagate with very low loss and long-lived coherence in a wide range of media \cite{Nat.Pho2009-3-687}. Strong interaction between matter and propagating photons is critical for quantum networks. A common approach to realizing strong matter-photons coupling is to use quantum emitters coupled to one-dimensional photonic waveguides. Recently, the photons transport in one dimensional waveguide has been studied extensively both theoretically \cite{OL2005-30-2001, PRL2007-98-153003, NJP2010-12-043052, PRA2010-82-063821, PRA2011-84-063832, PRL2011-106-053601, PRL2013-110-113601, PRL2013-111-063601, PRL2012-109-253603, PRA2013-88-043806, JOSAB2013-30-978, JOSAB2013-30-1135, PRL2014-113-243601} and experimentally \cite{Sci2010-327-840, Sci2013-342-1494}.

Quantum router \cite{PRA2012-85-021801, PRX2013-3-031013, PRA2013-87-062333, PRL2013-111-103604, PRA2014-89-013805, SR2014-4-4820, arxiv2014-6509, OE2015-23-22955, PRL2009-102-083601, PRL2011-107-073601, PRA2011-83-043814, Sci2014-345-903} for controlling the path of signals is an essential element for the quantum network. Recently, many theoretical \cite{PRA2012-85-021801, PRX2013-3-031013, PRA2013-87-062333, PRL2013-111-103604, PRA2014-89-013805, SR2014-4-4820, arxiv2014-6509, OE2015-23-22955} and experimental\cite{PRL2009-102-083601, PRL2011-107-073601, PRA2011-83-043814, Sci2014-345-903} researches on quantum routing of single photon have been proposed in various systems. Zhou and Lu et al. proposed schemes for the quantum routing of single photon with two output channels by a three-level atom with the help of classical field \cite{PRL2013-111-103604, PRA2014-89-013805}. A single-photon router with multiple input and output ports has been studied by Yan et al. using the interferences related to the phase differences of photons \cite{arxiv2014-6509}. In these configurations, the single photon injected into the input channel can be redirected into another channel with a maximum probability no more than $1/2$ due to the emission symmetry of quantum emitter. Thus a multichannel quantum router, which can be redirected into either of the output channels with an extremely high probability, will be of considerable interest.

In this paper, we propose a simple and scalable scheme for the quantum routing of single photon with two output channels, which is realized by two transmission lines (TLs). Here the active element of the router is a system consisting of two artificial atoms, which are two superconducting transmon qubits strongly coupled to two TLs. Our scheme is based on quantum interferences between the scattered photons by two atoms. An adequate engineering of the system can be used to break the emission symmetry of two atoms, channeling the emitted photons preferentially into one of the two TLs. Under standing wave condition and suitable choice of system parameters, we find that the reflectance and the backward transfer rate of single photon are zero due to quantum destructive interference. In asymmetric reflected phase shift case, single photon from the incident channel can be perfectly transmitted into another channel due to quantum constructive interference. In symmetric reflected phase shift case, the quantum interferences redirect the input photon into the other channel completely when input photon near resonantly interact with the atoms at the same decay rates. These are in sharp contrast to the previous theoretical schemes \cite{PRL2013-111-103604, PRA2014-89-013805,arxiv2014-6509} concentrating on single-atom case. Analytical and numerical results show that the single photon router can be controlled by adjusting the detuning of the atoms and interatomic distance, without any classical field.

\section{The Model}

\begin{figure}[t]
\centering
\includegraphics[width=\linewidth]{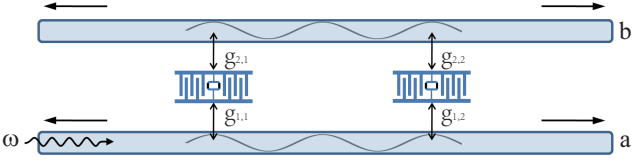}
\caption{(Color online) Schematic diagram of the quantum routing of the single photon. The two transmission lines are connected by two distant transmon qubits (separated by L). An input photon from the left side of TL-a will be transmitted, reflected or transferred to TL-b.}
\label{F1}
\end{figure}

As shown in Fig. \ref{F1}, our model consists of two superconducting transmon qubits separated by a distance $L$ coupled to two one-dimensional TLs. The transmission lines have a continuous spectrum of photonic modes propagated to left and right. Under the dipole and rotating-wave approximation, the Hamiltonian for whole system in interaction picture reads ($\hbar=1$)\cite{PRA2010-82-063821}
\begin{equation}
\begin{split}
H&=\sum_{j=a,b} \int d\omega \omega (r^{\dag}_{\omega j}r_{\omega j}+l^{\dag}_{\omega j}l_{\omega j})+\sum_{n=1,2}\omega_n\sigma^+_n\sigma^-_n 
\\&+\sum_{j=a,b}\sum_{n=1,2} \Big[\int d\omega g_{n,j}\sigma^+_n r_{\omega j}e^{i\omega z_n/v_g}
\\&+\int d\omega g_{n,j}\sigma^+_n l_{\omega j}e^{-i\omega z_n/v_g}\Big]+H.c.
\label{e1}
\end{split}
\end{equation}
where $r_{\omega j}$ ($l_{\omega j}$) is annihilation operator for a
right(left)-propagating photon with frequency $\omega$ in the TL-j.
$\sigma^+_1$ ($\sigma^+_2$) and $\sigma^-_1$ ($\sigma^-_2$) are
raising and lowering operators for the qubit-1 (qubit-2) with
transition frequency $\omega_1$ ($\omega_2$), at position $z_1=-L/2$
($z_2=L/2$), respectively. $v_g$ is group velocity of the photons.
$g_{n,j}$ is coupling strength of the qubit-n to the TL-j. Here we
assume that the frequency width of pulses is much smaller than
carrier frequency, we thus take the coupling strength $g_{n,j}$ to
be independent of the photon frequency\cite{PRL2013-111-063601}.

\section{Single-photon quantum router}

In this paper, we are interested in the single photon scattering by two separated qubits, i.e. the two qubits act as a quantum router. We introduce the input-output formalism for few-photon transport based on scattering theory \cite{PRA2010-82-063821,PRA2011-84-063832}. Then we define the input and output field operators that travel to the right as $o_{in,j}(t)=\frac{1}{\sqrt{2\pi}}\int d\omega e^{-i\omega(t-t_0)}o_{\omega j}(t_0)$, and $o_{out,j}(t)=\frac{1}{\sqrt{2\pi}}\int d\omega e^{-i\omega(t-t_1)}o_{\omega j}(t_1)$ ($o=r, l$), in the limit $t_0\rightarrow -\infty$ and $t_1\rightarrow \infty$. Following the procedure in Refs. \cite{PRA2010-82-063821, PRA2011-84-063832}, we obtain the input-output formalism equations as
\begin{equation}
\begin{split}
r_{out,j}(t)&=r_{in,j}(t)-i\sqrt{2\pi}g_{1,j}\sigma^-_1(t-\frac{L}{2v_g})
\\&-i\sqrt{2\pi}g_{2,j}\sigma^-_2(t+\frac{L}{2v_g}),
\label{e2}
\end{split}
\end{equation}
\begin{equation}
\begin{split}
l_{out,j}(t)&=l_{in,j}(t)-i\sqrt{2\pi}g_{1,j}\sigma^-_1(t+\frac{L}{2v_g})
\\&-i\sqrt{2\pi}g_{2,j}\sigma^-_2(t-\frac{L}{2v_g}),
\label{e3}
\end{split}
\end{equation}
\begin{equation}
\begin{split}
\dot{\sigma}^-_1(t)&=(-i\omega_1-\sum_{j=1,2}2\pi g^2_{1,j})\sigma^-_1(t)
\\&+\sum_{j=1,2}i\sqrt{2\pi}g_{1,j}\sigma^z_1(t) [r_{in,j}(t+\frac{L}{2v_g})
\\&+l_{in,j}(t-\frac{L}{2v_g})-\sqrt{2\pi}g_{2,j}\sigma^-_2(t-\frac{L}{v_g})],
\label{e4}
\end{split}
\end{equation}
\begin{equation}
\begin{split}
\dot{\sigma}^-_2(t)&=(-i\omega_1-\sum_{j=1,2}2\pi g^2_{2,j})\sigma^-_2(t)
\\&+\sum_{j=1,2}i\sqrt{2\pi}g_{2,j}\sigma^z_2(t) [r_{in,j}(t-\frac{L}{2v_g})
\\&+l_{in,j}(t+\frac{L}{2v_g})-\sqrt{2\pi}g_{1,j}\sigma^-_1(t-\frac{L}{v_g})].
\label{e5}
\end{split}
\end{equation}

According to the input-output relation, the output states of our device for given input states are governed by the scattering matrix (S-matrix), $|\Psi_{out}\rangle=S|\Psi_{in}\rangle$. We assume that, initially, a photon is injected into the TL-a from the left side, as shown in Fig. \ref{F1}. The S-matrix elements describe the single-photon scattering amplitudes as $S_{o,j}(k,p)=\langle 0|o_{out,j}(p)r_{in,1}^{\dag}(k)|0\rangle$. Here the $k$ and $p$ refer to free-photon frequency incidently and outgoingly, respectively.

Fourier transforming of the Eqs. \ref{e2} and \ref{e3} leads to the single-photon scattering amplitude,
\begin{equation}
\begin{split}
S_{r,j}(k,p)&=\langle 0|r_{in,j}(p)|k^+\rangle
\\&-i\sqrt{2\pi}g_{1,j}\mathcal {F}^{-1}[\langle 0|\sigma^-_1(t-\frac{L}{2v_g})|k^+\rangle]
\\&-i\sqrt{2\pi}g_{2,j}\mathcal {F}^{-1}[\langle 0|\sigma^-_2(t+\frac{L}{2v_g})|k^+\rangle],
\label{e6}
\end{split}
\end{equation}
\begin{equation}
\begin{split}
S_{l,j}(k,p)&=-i\sqrt{2\pi}g_{1,j}\mathcal {F}^{-1}[\langle 0|\sigma^-_1(t+\frac{L}{2v_g})|k^+\rangle]
\\&-i\sqrt{2\pi}g_{2,j}\mathcal {F}^{-1}[\langle 0|\sigma^-_2(t-\frac{L}{2v_g})|k^+\rangle],
\label{e7}
\end{split}
\end{equation}
where $|k^+\rangle\equiv r_{in,1}^{\dag}(k)|0\rangle$, $\mathcal {F}^{-1}[]$ denotes the inverse Fourier transform operator. In order to calculate $\mathcal {F}^{-1}[\langle 0|\sigma^-_{(1,2)}(t)|k^+\rangle]$, we use Eqs. \ref{e4} and \ref{e5} to perform an inverse Fourier transformation, so that we can derive iterative formulas of $\mathcal {F}^{-1}[\langle 0|\sigma^-_{(1,2)}(t)|k^+\rangle]$,
\begin{equation}
\begin{split}
\mathcal {F}^{-1}[\langle 0|\sigma^-_{1}(t)|k^+\rangle]&=iS_1\mathcal {F}^{-1}[\langle 0|r_{(in,1)}(t+\frac{L}{2v_g})|k^+\rangle]
\\&+T_1\mathcal {F}^{-1}[\langle 0|\sigma^-_{2}(t-\frac{L}{v_g})|k^+\rangle],
\label{e8}
\end{split}
\end{equation}
\begin{equation}
\begin{split}
\mathcal {F}^{-1}[\langle 0|\sigma^-_{2}(t)|k^+\rangle]&=iS_2\mathcal {F}^{-1}[\langle 0|r_{(in,1)}(t-\frac{L}{2v_g})|k^+\rangle]
\\&+T_2\mathcal {F}^{-1}[\langle 0|\sigma^-_{1}(t-\frac{L}{v_g})|k^+\rangle],
\label{e9}
\end{split}
\end{equation}
where
\begin{equation}\nonumber
\begin{split}
S_n&=\frac{-\sqrt{\gamma_{n,a}}}{\gamma_{n,a}+\gamma_{n,b}}\cos\theta_n e^{i\theta_n}, \\
T_n&=\frac{-\sqrt{\gamma_{1,a}\gamma_{2,a}}-\sqrt{\gamma_{1,b}\gamma_{2,b}}}{\gamma_{n,a}+\gamma_{n,b}}\cos\theta_n e^{i\theta_n},
\end{split}
\end{equation}
$\gamma_{n,j}=2\pi g^2_{n,j}$ is the decay rate from the qubit-n to the TL-j, and
$\theta_n=\arctan[(p-\omega_n)/(\gamma_{n,a}+\gamma_{n,b})]$ is the phase shift given by the qubit-n to the light upon reflection \cite{PRL2013-110-113601}. We plug iteration equations \ref{e8} and \ref{e9} into Eqs. \ref{e6} and \ref{e7} to obtain the single-photon scattering matrix $S_{o,j}(k,p)=t_o^j\delta(p-k)$, where
\begin{widetext}
\begin{equation}
\begin{split}
t^a_r&=1+\frac{\sqrt{\gamma_{1,a}}S_1+\sqrt{\gamma_{1,a}}T_1 S_2 e^{ik2L/v_g}+\sqrt{\gamma_{2,a}}S_2+\sqrt{\gamma_{2,a}}T_2 S_1}{1-T_1 T_2 e^{ik2L/v_g}}, \\
t^b_r&=\frac{\sqrt{\gamma_{1,b}}S_1+\sqrt{\gamma_{1,b}}T_1 S_2 e^{ik2L/v_g}+\sqrt{\gamma_{2,b}}S_2+\sqrt{\gamma_{2,b}}T_2 S_1}{1-T_1 T_2 e^{ik2L/v_g}}, \\
t^a_l&=\frac{\sqrt{\gamma_{1,a}}S_1+(\sqrt{\gamma_{1,a}}T_1 S_2+\sqrt{\gamma_{2,a}}S_2+\sqrt{\gamma_{2,a}}T_2 S_1) e^{ik2L/v_g}}{1-T_1 T_2 e^{ik2L/v_g}}, \\
t^b_l&=\frac{\sqrt{\gamma_{1,b}}S_1+(\sqrt{\gamma_{1,b}}T_1 S_2+\sqrt{\gamma_{2,b}}S_2+\sqrt{\gamma_{2,b}}T_2 S_1) e^{ik2L/v_g}}{1-T_1 T_2 e^{ik2L/v_g}},
\label{e10}
\end{split}
\end{equation}
\end{widetext}
$t^a_r$ ($t^a_l$) is the transmitted (reflected) amplitude and $t^b_r$ ($t^b_l$) is the forward (backward) transfer amplitude. where we have used the equation \cite{PRA2010-82-063821},
\begin{equation}
\mathcal {F}^{-1}[\langle 0|r_{in,1}(t-\frac{2NL}{v_g})|k^+\rangle]=e^{ik\frac{2NL}{v_g}}\delta(p-k).
\label{e11}
\end{equation}
According to the probability conservation, the scattering amplitudes satisfy $|t^a_r|^2+|t^a_l|^2+|t^b_r|^2+|t^b_l|^2=1$. It should be noted that single-photon scattering amplitude depends on the interatomic distance through the dephasing term $e^{ik2L/v_g}$, so we can adjust the interatomic distance to make the incoming photon in TL-a transfer to TL-b. It is the two distant atoms that implements the photon-redirection function of the quantum router.

To see the effect of the single-photon router more clearly, we look
into the situation in which the distance between the qubits
satisfies the standing wave condition (i.e.
$2kL/v_g+\theta_1+\theta_2=2n\pi$). If we set
$\cos\theta_1=\cos\theta_2$ (i.e. $\theta_1=\pm\theta_2$), and
$\gamma_{1,a}/\gamma_{1,b}=\gamma_{2,a}/\gamma_{2,b}$, we find that
the reflectance $R^a=|t^a_l|^2$ and the transfer rate
$T^b_l=|t^b_l|^2$ are zero. This is because of the quantum
interferences. The scattering amplitude $t^a_l$ ($t^b_l$) vanish
when left-moving photons scattered by the two qubits interfere
destructively and cancel each other completely in TL-a (TL-b).

We first consider the situation in which reflected phase shift given
by the qubit-1 is equal to the one given by the qubit-2, i.e.
$\theta_1=\theta_2$. In this case, the reflectance $R^a$ and the
transfer rate $T^b_l$ vanish and  the conservation relation changes
into $|t^a_r|^2+|t^b_r|^2=1$. Inserting $\theta_1=\theta_2$ and
$\gamma_{1,a}/\gamma_{1,b}=\gamma_{2,a}/\gamma_{2,b}$ into Eqs.
\ref{e10}, we obtain the single-photon scattering probability as
below:
\begin{equation}
\begin{split}
T^a&=|t^a_r|^2=1-\frac{4\gamma_{1,a}\gamma_{1,b}}{(\gamma_{1,a}+\gamma_{1,b})^2}\cos^2\theta, \\
T^b_r&=|t^b_r|^2=\frac{4\gamma_{1,a}\gamma_{1,b}}{(\gamma_{1,a}+\gamma_{1,b})^2}\cos^2\theta.
\label{e12}
\end{split}
\end{equation}
Obviously, only if the input photon near resonantly interact with
the qubits and the decay rate $\gamma_{1,a}$ is equal to
$\gamma_{1,b}$, i.e. $\cos\theta\rightarrow 1$ and
$\beta=\gamma_{1,a}/\gamma_{1,b}=1$, the input photon can be
redirected into the other output channel completely. This is in
sharp contrast to the single-atom case, where perfect transfer rate
cannot be observed \cite{PRL2013-111-103604,PRA2014-89-013805}. This
is the result of quantum constructive interference. When the
reflected phase shift $\cos\theta\rightarrow 0$, single photon can
not travel in TL-b, i.e. single photon is totally transmitted to the
right side of TL-a. The condition $\cos\theta\rightarrow 0$
corresponds to the large detuning case. When
$\gamma_{1,a}\ll\gamma_{1,b}$ or $\gamma_{1,a}\gg\gamma_{1,b}$, we
can also get the result $T^a\approx 1$, which is similar to the
large detuning case. In this case, one of two TLs is decoupled to
two qubits. It means that two TLs are not connected, and the photon
will be transmitted completely.

\begin{figure}[tbp]
\centering
\subfigure[]{\label{F2a}\includegraphics[width=0.49\linewidth]{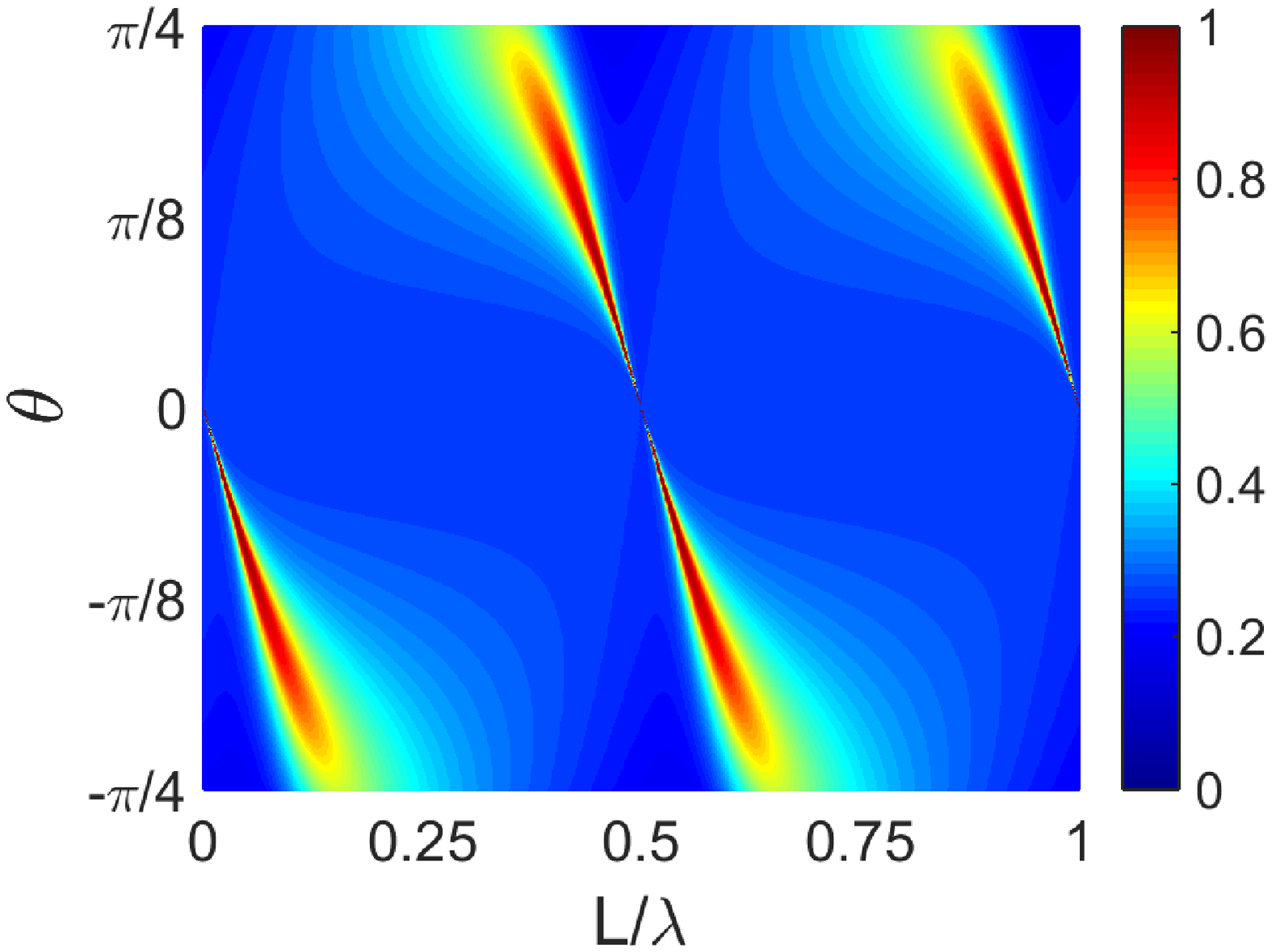}}
\subfigure[]{\label{F2b}\includegraphics[width=0.49\linewidth]{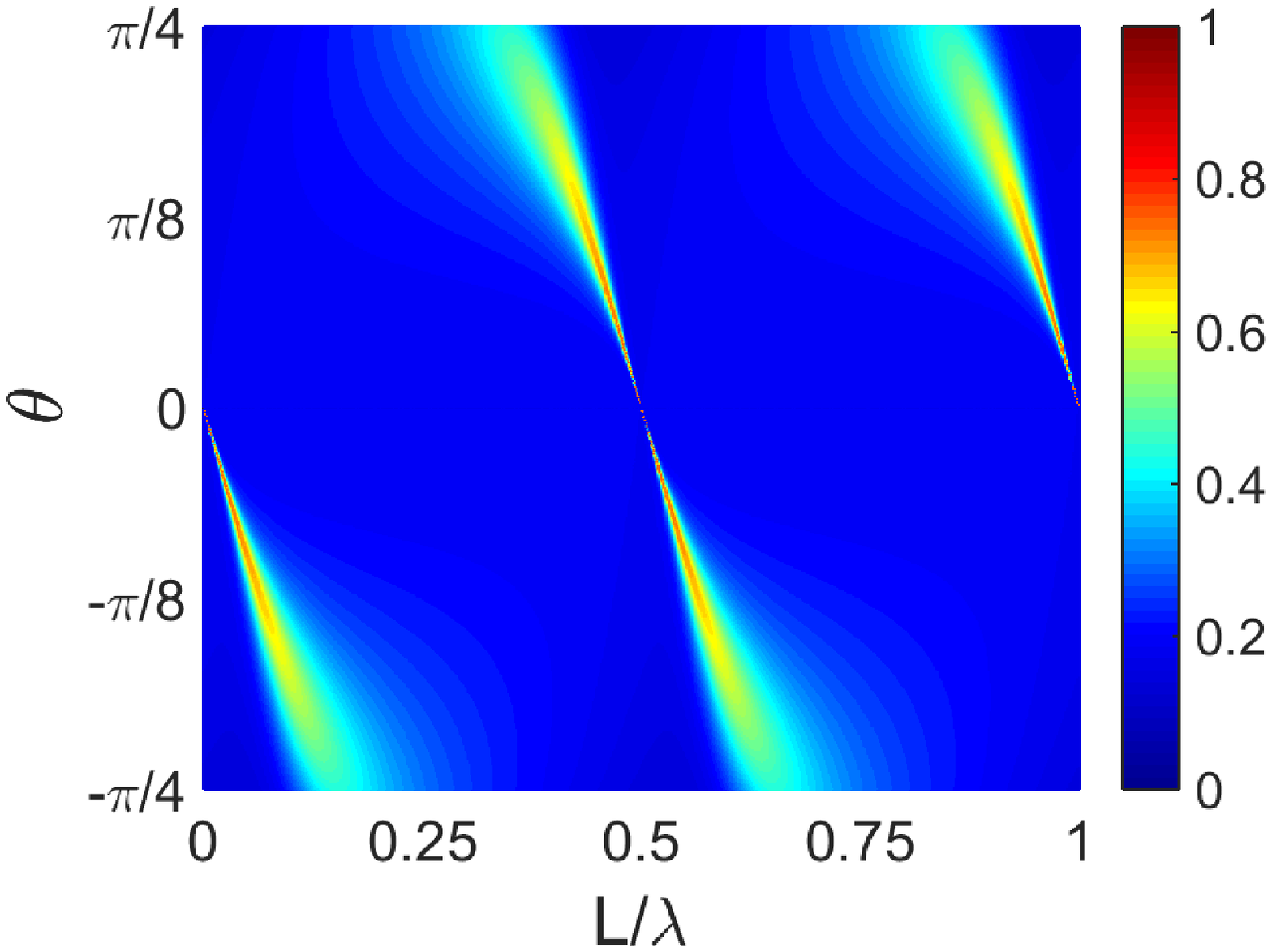}}
\caption{(Color online) Color maps of the transfer rate $T^b_r$  as a function of the phase shift $\theta=\theta_1=\theta_2$ and distance $L$. (a) $\beta$=1, (b) $\beta$=3.}
\label{F2}
\end{figure}

\begin{figure}[tbp]
\centering
\subfigure[]{\label{F3a}\includegraphics[width=0.95\linewidth]{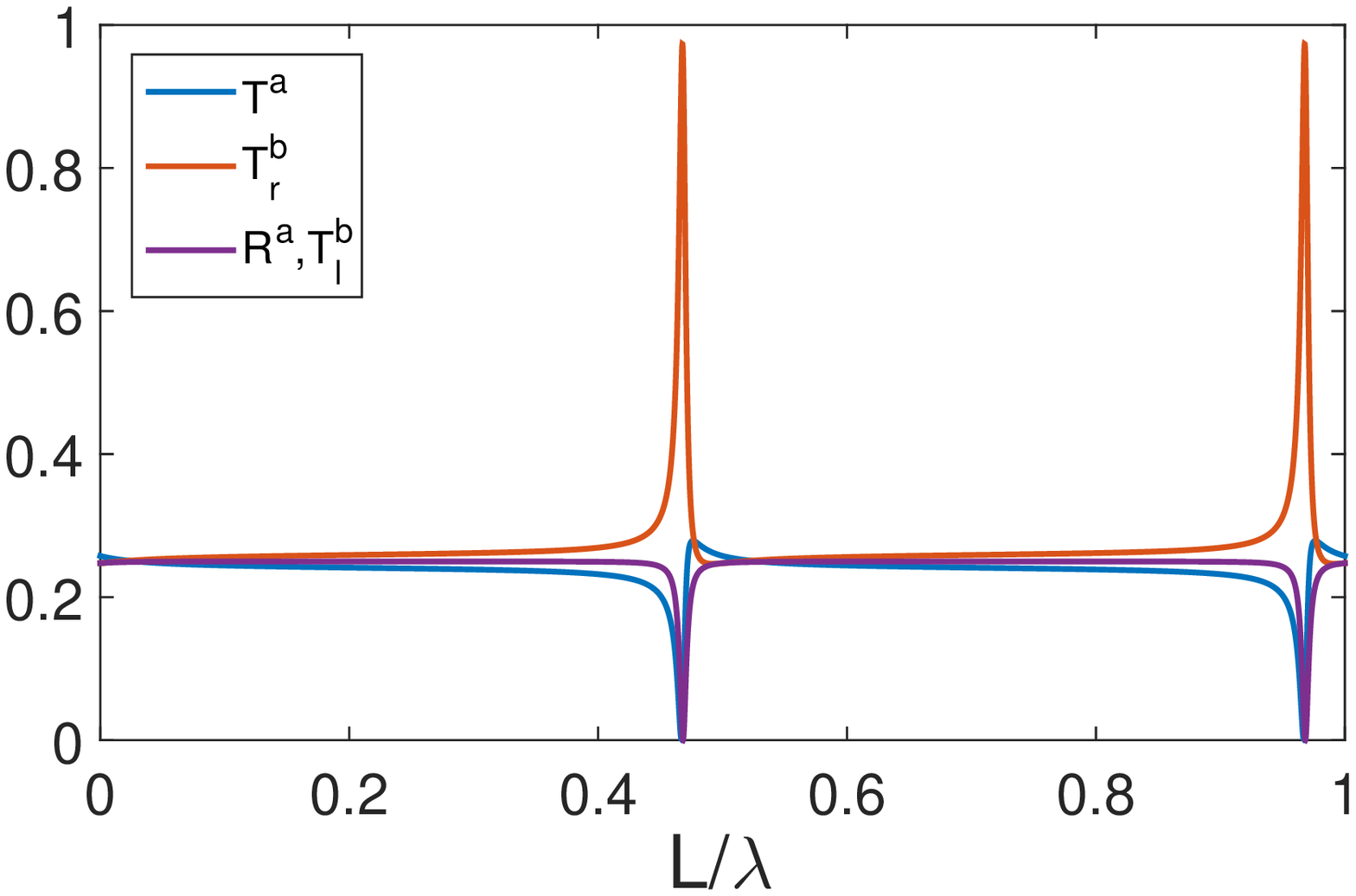}}
\subfigure[]{\label{F3b}\includegraphics[width=0.95\linewidth]{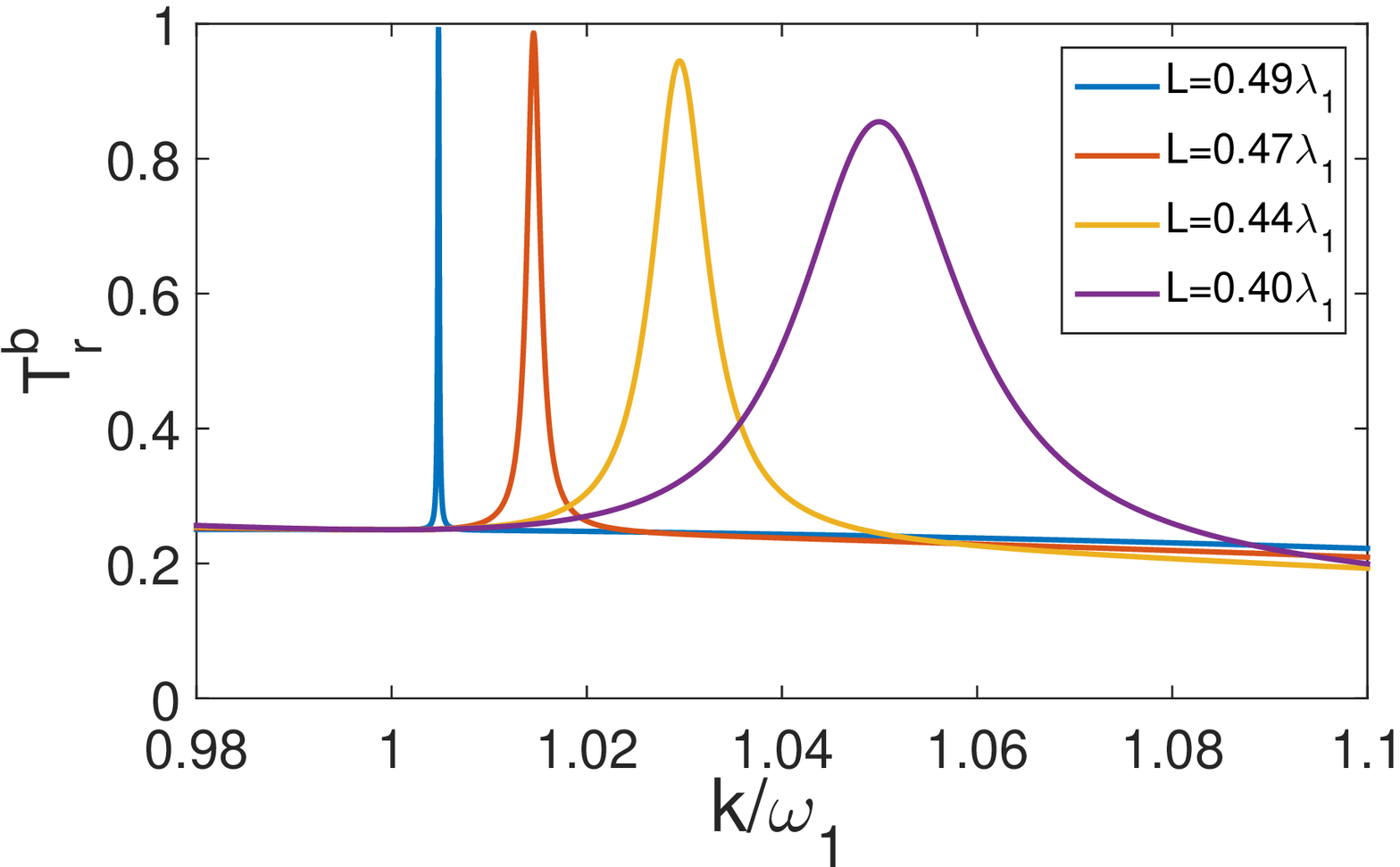}}
\caption{(Color online) (a) the single-photon scattering probabilities as a function of the distance $L$ in symmetric reflected phase shift case. We set $\theta=0.2$. (b) the transfer rate $T^b_r$ as a function of the wave number $k$ for various values of $L$. We set $\lambda_1=2\pi v_g/\omega_1$. Both curves are plotted for $\omega_1=\omega_2=20$.}
\label{F3}
\end{figure}

To show the details of the quantum routing of the single photon in the $\theta_1=\theta_2$ case, we plot the transfer rate $T^b_r$ against the phase shift and distance in Fig. \ref{F2}. For simplicity, in the following we will consider $\gamma _{1,a}=\gamma_{2,a}=1$ and $L$ in units of the photon wavelength $\lambda=2\pi v_g/k$. We notice that there is a large window of perfect transfer rate: $T^b_r\approx 1$, when the decay rates from qubits to TLs are equal. When the decay rates are not equal, input photon can not be redirected into the other channel completely. The single-photon scattering probabilities can be controlled by adjusting the detuning and interatomic distance. Fig. \ref{F3a} shows the single-photon scattering probability against interatomic distance $L$. We find that the scattering probability is a periodic function
about distance $L$. When the distance between the qubits satisfies
standing wave condition, the transfer rate $T^b_r$ is approximately
unity. If the distance can not satisfy standing wave condition, the
probabilities of single photon redirected in the four output ports
are approximately equal. This is because reflections from the two
qubits can not interfere destructively and cancel each other
completely. As shown in Fig. \ref{F3b}, we plot the transfer rate $T^b_r$
as a function of the wave number $k$ for various values of $L$. The
width of this transfer window sees a dramatic increase along with
the interatomic distance. On the contrary, we find that the
altitudes of the transfer rate window decrease with the distance.

Next we consider the case that the reflected phase shift from the qubit-1 is equal to negative one from the qubit-2 (i.e. $\theta_1=-\theta_2$). In this case, the standing wave condition changes into $2kL/v_g=2n\pi$. Different from symmetric reflected phase shift case, we find that the right-moving photons in TL-b from the two qubits interfere destructively and cancel each other completely due to antisymmetric reflected phase shift, i.e. single photon are completely transmitted.

\begin{figure}[tbp]
\centering 
\subfigure[]{\label{F4a}\includegraphics[width=0.49\linewidth]{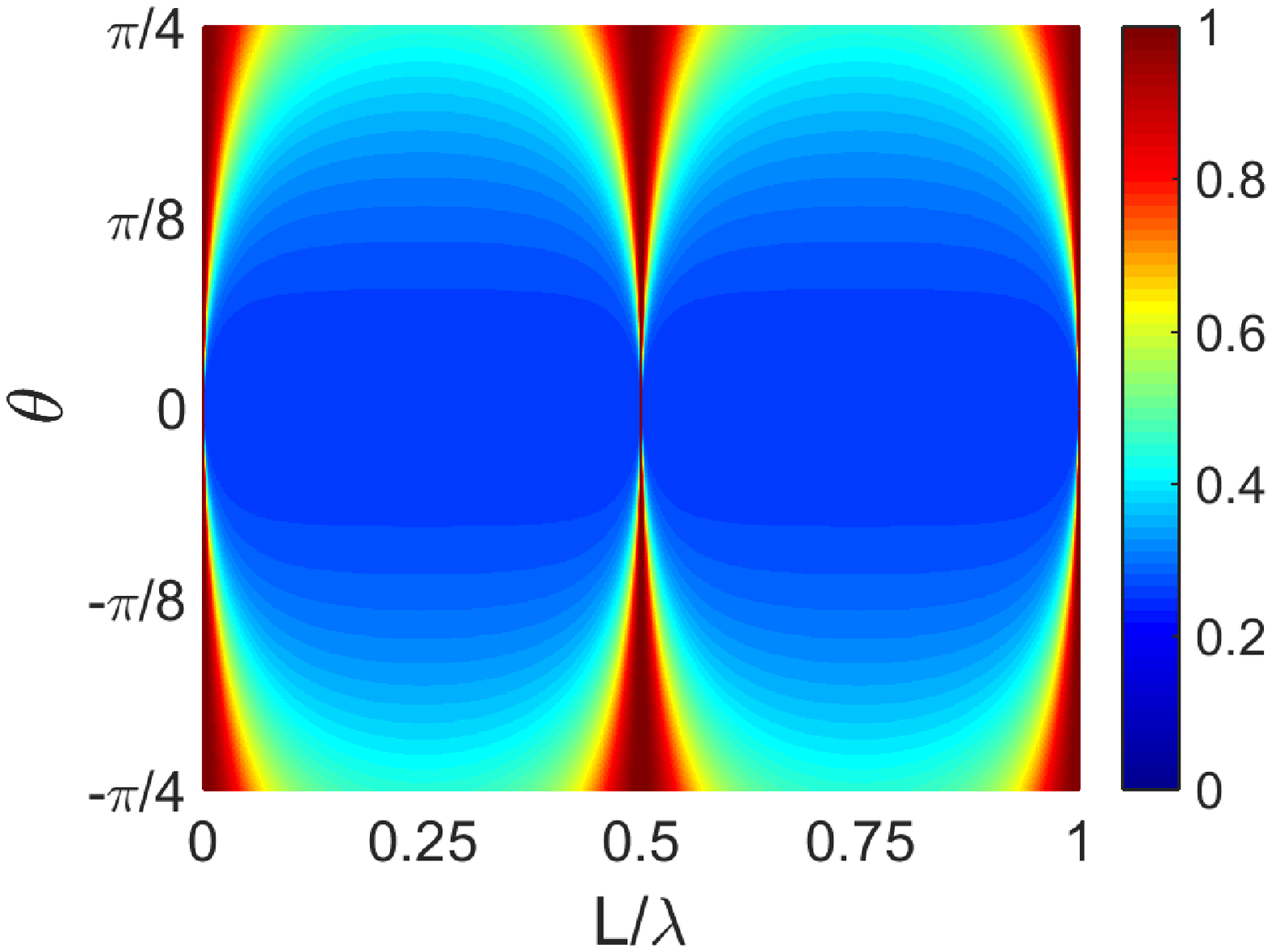}}
\subfigure[]{\label{F4b}\includegraphics[width=0.49\linewidth]{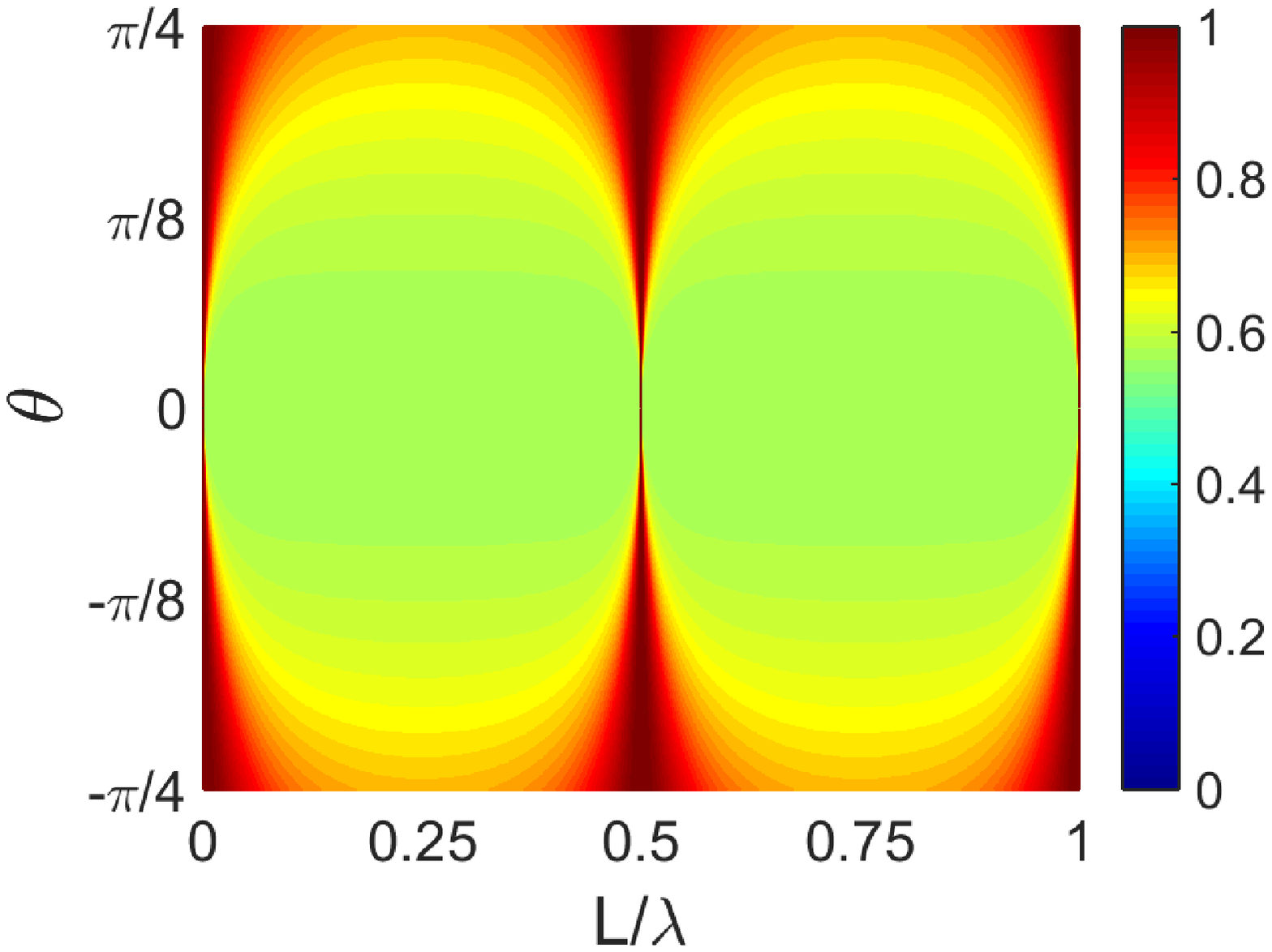}}
\caption{(Color online) Color maps of the transmission $T^a$  as a function of the phase shift $\theta=\theta_1=-\theta_2$ and distance $L$. (a) $\beta$=1, (b) $\beta$=3.}
\label{F4}
\end{figure}

\begin{figure}[tbp]
\centering 
\subfigure[]{\label{F5a}\includegraphics[width=0.95\linewidth]{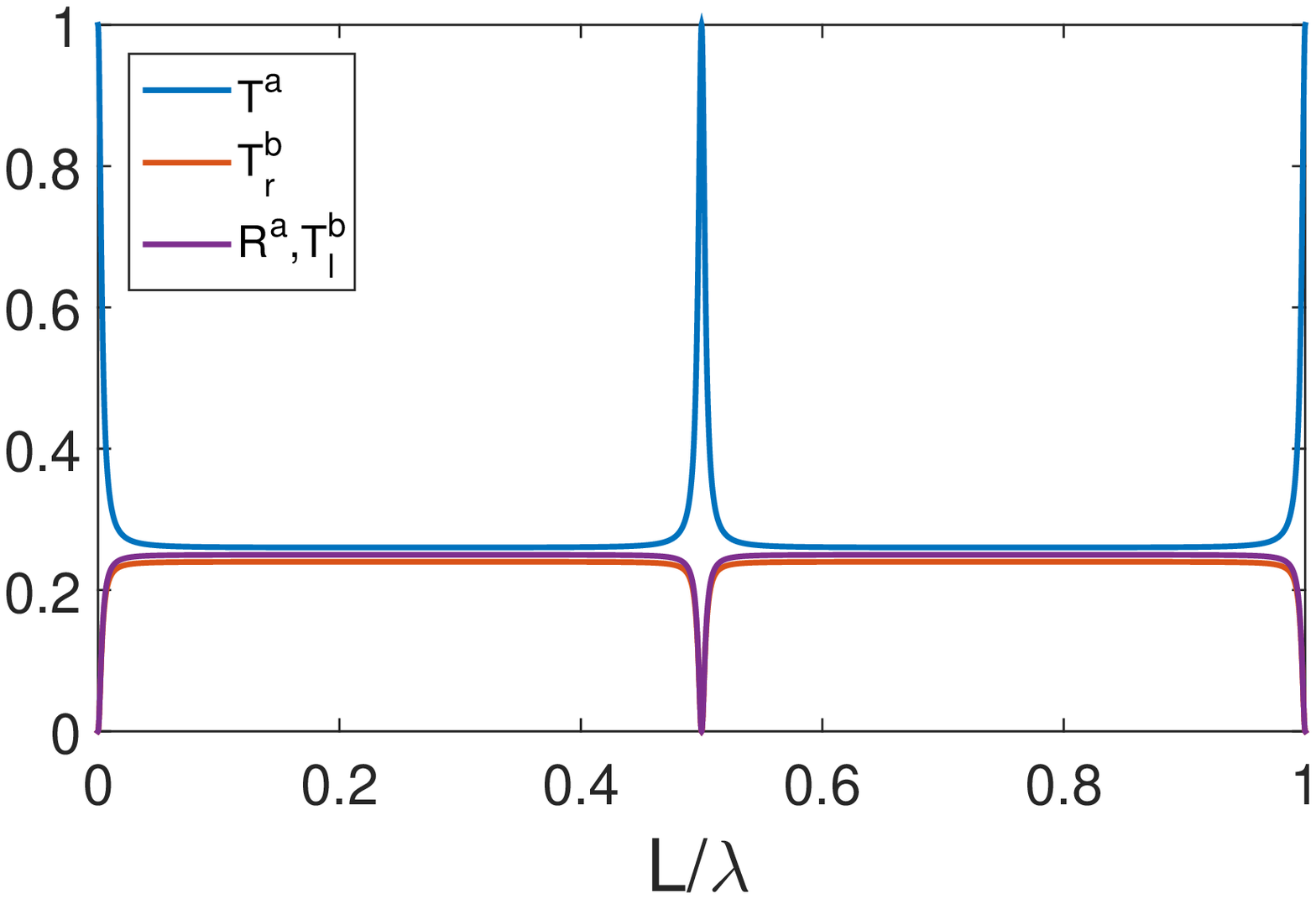}}
\subfigure[]{\label{F5b}\includegraphics[width=0.95\linewidth]{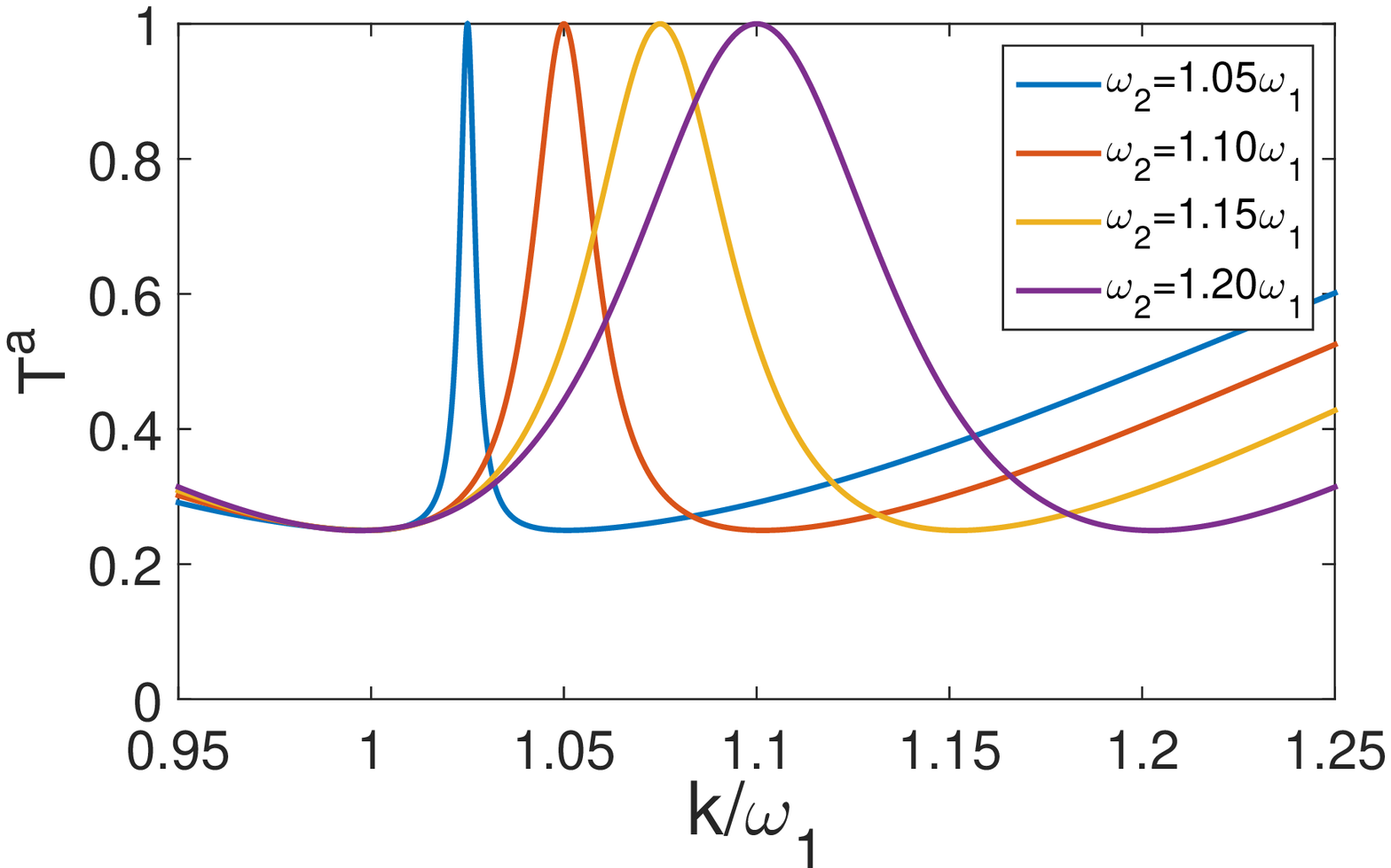}}
\caption{(Color online) (a) the single-photon scattering probabilities as a function of the distance $L$ in asymmetric reflected phase shift case.  We set $\theta=0.2$. (b) the transfer rate $T^a$ as a function of the wave number $k$ for various values of $\omega_2$. We set $L=4\pi v_g/(\omega_1+\omega_2)$.  Both curves are plotted for $\omega_1=20$.}
\label{F5}
\end{figure}

As shown in Fig. \ref{F4}, there is a large window of perfect transmission: $T^a_r\approx 1$, even when the detuning of single photon is within the resonance linewidth. This has obvious differences with the single-atom case, where perfect transmission is only possible for far off-resonance photons \cite{PRL2013-111-103604,PRA2014-89-013805}. Similar to symmetric reflected phase shift case, the scattering probability is a periodic function about distance $L$ as shown in Fig. \ref{F5a}. Under the standing wave condition, the transmission $T^a$ is approximately unity. Fig. \ref{F5b} shows the transfer rate $T^a$ as a function of the wave number $k$ for various values of $\omega_2$. The width of this central transparency window is only determined by the detuning between the incoming photon and the qubits. Furthermore, the system is transparent for $(k-\omega_2)\rightarrow\pm\infty$, where the incoming photon is far off-resonant and thus does not interact with the qubits. Unlike symmetric reflected phase shift case, we find the altitudes of the transmission window is constant.

\section{Conclusion}

In conclusion, we have presented a detailed investigation on the routing of single photon by two distant artificial atoms. The routing of photons can be achieved by reflected phase shift given by the atoms and distance between the atoms. Unlike the single-atom case, the reflectance and the backward transfer rate of single photon vanish under standing wave condition and suitable choice of system parameters. We have shown that single photon incident from the left sides of TL-a can be totally transfer to the right side of TL-b at $\cos\theta_1=\cos\theta_2\rightarrow 1$ when the decay rates of atoms are equal. For the asymmetric reflected phase shift case, the perfect transmission occurs even when the detuning of single photon is within the resonance linewidth. We believe that our device can be readily implemented to the existing technology.

\section{Acknowledement}

We acknowledge grant support from the NSF No.11375036,  No.11505024
and No.11574041 and the Fundamental Research Funds for the Central
Universities No.DUT10LK10.

\end{document}